# Investigations of Attractor Behavior over the Decay of Modular RBNs

Shaun Deaton[1] & Seth Frey[2]

[1] Informatics, Indiana University, 901 E. 10th Street, Bloomington, Indiana 47408. srdeaton@indiana.edu
[2] Cognitive Science, Indiana University, 834 Eigenmann hall, 1900 E. 10th Street, Bloomington, Indiana 47408. sethfrey@indiana.edu

send correspondence to sethfrey@indiana.edu

**Abstract.** When is it safe to approximate a complicated random Boolean network (RBN) as a simplified, easier to model RBN? When can static measures of network structure be reliably used to infer the network's dynamics? This simple experiment tests the ability of disjoint modular RBNs to approximate the dynamics of progressively more interconnected RBNs, while characterizing the performance of both static and dynamic measures of modularity as both break down. We find that, at least in the small networks investigated, the Newman 2004 [1] measure of static modularity performs as well as a more complex dynamic measure of modularity, and that the progressively increasing failure of one tracks that of the other. The dynamic measure is based on the Hamming distance of attractor schemata in rewired networks from those in perfectly modular networks. This result holds for a range of p-values.

**Introduction**

As models of biological processes, Boolean networks omit enough important empirical phenomena to make a biologist blush. Yet they still provide valuable insight into biological processes. For example, in [2], the authors reduced a model of a Drosophila segment polarity network by isolating a collection of nodes representing a single cell in a multi-cellular design. When distilling a complex system down to a model, what simplifications are safe? We investigate this question in the modeling of modularity. Insofar as multi-scale systems are modular systems, it is important to understand what simplifying assumptions are safe. For example, is a completely modular system a good model of a 'loosely coupled' modular system? And at what rate do we lose fidelity when modeling the dynamics of a tightly interconnected system with a simpler loosely coupled system?

We compare both the topology and attractor structure of ideal modular networks to networks whose modularity is decaying under a regime of random rewiring. We then measure the decay of the fidelity of the former as a model of the latter. We find that, in approximating both the topology and dynamics of the more interconnected system, there exist static and dynamics measures that are qualitatively interchangeable. Specifically, Newman's leading eigenvector measure, a static modularity metric, has an (empirical) linear relation to a more complex dynamic measure of attractor structure, and is therefore no less effective at predicting the fidelity of ideal modules as dynamic models of "loosely-coupled" networks. This affirms the power of topology to constrain network dynamics in modular systems.

**Methodology & Setup**

In this experiment, a network of 6 modules is composed of 6 disjoint NK graphs. Though we varied the number of modules from 2 to 12, and the value N for each module from 2 to 12, most of this work is with six modules of N=8. Though we explored other values of K, almost all work was with K=2. Csárdi's igraph package, via Python, provided visualization and an implementation of the Newman metric [3]. Though this ideally modular graph is conceptually only one graph, it is functionally six independent subgraphs. We sampled the attractors of this system by dissecting and compressing the large attractors of 8x6 bits into 6 subattractors of 8. These populations of subattractors formed the 'models' against which we compared the dynamics of the corresponding nodes of each module after the modularity has succumbed to a long regime of random rewiring.

After sampling for the attractors of the completely disjoint 'model' system, we randomly rewired the network, resampled the attractor space, and compared the original subattractors to the new subattractors.

Rewiring was performed in such a way as to preserve the in-degree, and therefore the Boolean function, of each node. This involved randomly selecting a node, randomly selecting one of its inputs and changing which node in the graph is providing the input. With 6 modules of equal size, there is only a 1/6 chance that the new input will be from the same module.

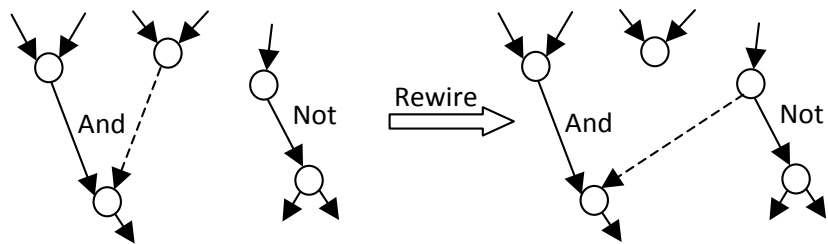

Figure 1: The dashed link is rewired by attaching its tail to different node.

A pair of attractors was compared by measuring the normalized Hamming distance between their respective representations as probabilistic schemata [4]. By this approach, a 2-cycle of three nodes: ((1,1,0),(1,0,0)), would be compressed into the schema (1,.5,0). The Hamming distance between two probabilistic schemata is the sum across each bit of the absolute values of the differences between the two values at that bit. We then normalized this value by the length of the bit string. For example, the normalized probabilistic Hamming distance between schemata (1,.5,0) and (0.5,0.5,0.5) would be 1/3.

In cases where the ideal module, and the new subattractor both had multiple attractors, all Hamming distances were measured and the minimum distance was taken as the model's best attempt at accounting for 'observed' dynamics in that section of the rewired graph.

Our rewiring mechanism preserved much of the initial seed RBNs structure compared to a general rewiring scheme in which a link is removed and replaced by a new link between any two nodes. The preservation of the original structure may be seen in our dynamic measure tracking the Hamming distance between attractors.

This measure tended to find a maximum distance around .33 compared with the average distance of .5 that would be expected from a completely randomized graph. However, this may be an artifact of our choice to choose the smallest Hamming distance from all pairs of attractors between the model and rewired system. Despite this possibility, the space of RBNs attainable from the seed RBN via this rewiring scheme is constrained.

Having such a constraint on the rewiring mechanism helps the RBN to preserve much of its initial structure while the modules are merging. Generally, it can happen that everything is changing, the out-degree, in-degree, and Boolean functions. That is, these changes can alter which nodes influence each other and how these influences are combined at a node. Also, since only a link's tail is moved, there remains a connection back to the link's head node and that node's containing module. Though, this link's tail may later be moved by another random rewire, if the process continued long enough, then it can rearrange any new connections. A rearrangement is perhaps the best way to refer to this rewiring process.

Static modularity was measured using the leading eigenvalue method for determining the optimal splitting of graphically represented data into communities based on the underlying graph's link structure. This algorithm is primarily used for graph partitioning, but also, provides a numerical value representing the likelihood that there exists distinct, often weakly connected, communities of nodes. Large positive values are found when communities are more likely and negative values indicate no underlying communities; other than the entire graph. Thus, modularity, as measured by the community measure, extracts information about the RBN's underlying graphical structure.

**Experiment**

The perfectly modular RBNs where rewired 200 times or more and the attractors where determined after each rewiring.

Their normalized Hamming distances from the seed RBN's attractors was calculated and then averaged. The change in modularity was also calculated at each step, helping to track communal decay of the modules.

In one experiment, a seed RBN consisting of six separate modules each with six nodes and k=2, was subjected to 250 rewirings. This was repeated 50 times, each with a different seed RBN. The modularity and normalized Hamming distance where averaged over all 50 data sets. The results are plotted below in figure 2; where the horizontal axis represents the 250 rewirings. While the left vertical axis, which is inverted, corresponds to the modularity and the right vertical axis corresponds to the normalized Hamming distance. The red plot depicts the modularity, labeled as 'q', this notation comes from Newman; and the light blue plot depicts the normalized Hamming distance. The plot shows that as the modularity decreases the hamming distance increases, respective of their own measurement axis. This is not too surprising, since a decrease in modularity means that the disconnected seed RBN is becoming more connected, resembling a single component. As this happens, the attractors should be changing greatly, and thus move further away from the attractors of the seed RBN.

The plot appears as it does, due to a linear transformation that flipped, scaled, and translated the modularity data. Though Newman's q decays slightly faster than the metric based on attractor structure, they track very well, indicating that decay in Newman value correlates with increases in the distance between attractors. Further, the Newman graph never diverges beyond the line demarcating the standard deviation of the Hamming distance, which stabilizes around .09 in most experiments.

The Hamming distance measure always started with zero distance from the model network. After a very fast increase, the distance always stabilized at 0.3. While this is short of complete randomness, this difference may be attributable to the fact that, out of all attractors from the model that could be used to represent an observed attractor, we always chose the one that gave the lowest Hamming distance from a given attractor. The Newman measure usually started around .85 with the perfectly modular system, even when provided a suggestion to look for the actual number of modules. This value decayed quickly over the course of the experiment, stabilizing around a value of .3 for the most randomized, completely non-modular systems.

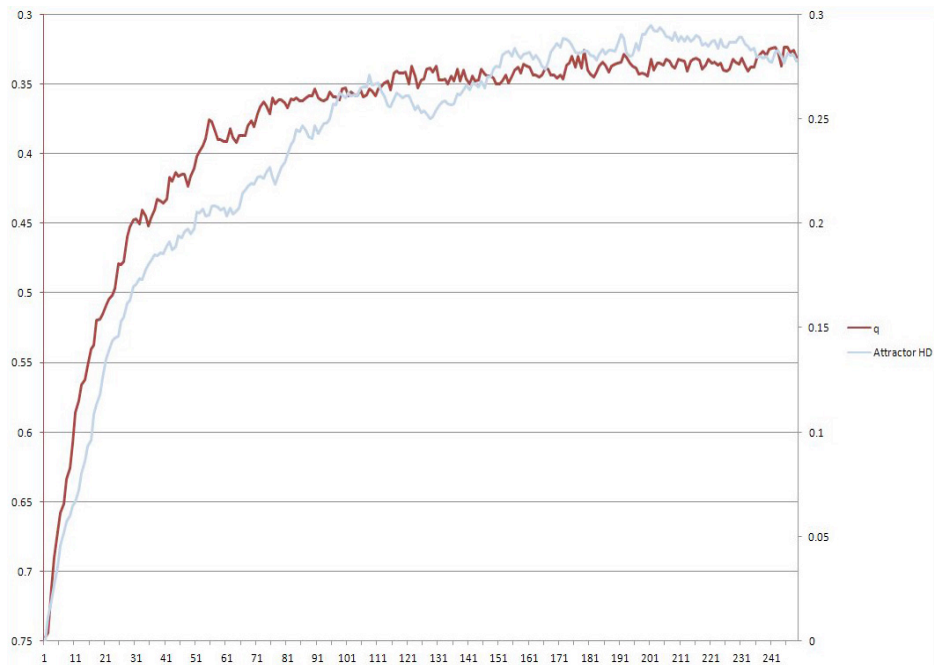

Figure 2: Plot of Hamming distance and linearly transformed modularity over 250 iterations. The left axis correspond to modularity q, the right to Hamming distance.

Conventional RBNs have Boolean functions with equal probability of being 1 or 0 given some random input [5]. Possibly because of their canalizing structure, models of biological networks often have a probability, p, reflecting the presence of functions like AND and OR, whose p are .75 and .25 respectively; to output a 0.  We tested our system on a range of p values, fixed to be the same at every node, and found that Newman and attractor Hamming distance continue to track well. This can be seen in figure 4 below, as compared with figure 3 above.

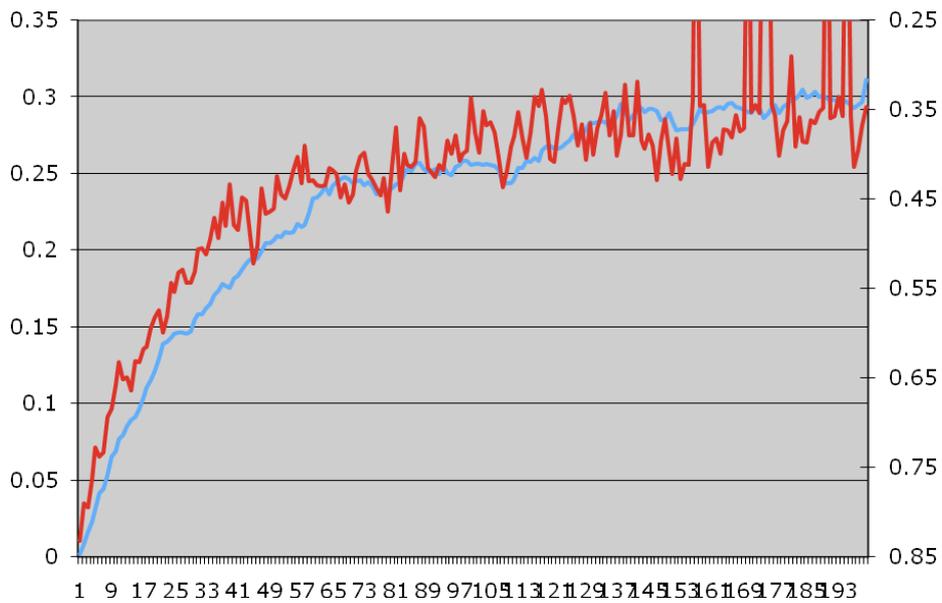

Figure3: Using a p=.5 value for the Boolean functions; Red is the modularity measured by the right axis, Light Blue is the average Hamming distance measured by the left axis.

## Conclusion & Future Work

Based on the data collected, some of which is shown above, there appears to be a linear relationship between communal decay in RBNs with disconnected components and how close, on average, the attractors are at each stage of the decay. Noting, of course, that the communal decay is restricted by the choice of rewiring method, and so greatly influences how the RBNs evolve. This affects how the attractors change, thus weighs upon the measured Hamming distance between sets of attractors.

We also observed that, by both measures, modularity decreased most dramatically after the first rewirings of the experiment. To answer one of our guiding questions, this suggests that the performance of an uncoupled system is only a good model of a coupled system for a very limited couplings.

In the networks examined, the fidelity of the models decreased too reliably to leave them any predictive power beyond the first few dozen rewirings.

Future work on this topic will include examining and checking the linear relation conjecture between communal decay and attractor trajectories of disconnected RBNs; under the given rewiring mechanism. Interestingly, we found that the scale parameter in the linear transformations performed of the modularity was around .6; finding out the source of this 60% rescaling is another phenomenon to address. Perhaps, also comparing the results to a general rewiring that can change all aspects of the seed RBN. Another direction is to address the concern with using an average over the Hamming distances between the attractor structures to determine how much the rewired RBN differs from the seed.